\documentclass[aps,pra,preprint,showpacs,superscriptaddress]{revtex4}
\usepackage{mathtools}
\usepackage{graphicx}

\begin{document}
\title{Spatio-temporal interference of photo electron wave packets and time scale of non-adiabatic transition in high-frequency regime}
\author{Koudai Toyota}
\email{koudai.toyota@cfel.de}
\affiliation{Center for Free-Electron Laser science, DESY, 22607 Hamburg, Germany}
\affiliation{Max Planck Institute for the Physics of Complex Systems, 01187 Dresden, Germany }
\begin{abstract}
The method of the envelope Hamiltonian 
[K. Toyota, U. Saalmann, and J. M. Rost, New J. Phys. {\bf 17}, 073005~(2015)] is applied
to further study a detachment dynamics of a model negative ion in one-dimension in high-frequency regime. 
This method is based on the Floquet approach, but the time-dependency of an envelope function 
is explicitly kept for arbitrary pulse durations. Therefore, it is capable of describing
not only a photo absorption/emission but also a non-adiabatic transition which is induced by the time-varying 
envelope of the pulse. It was shown that the envelope Hamiltonian accurately retrieves the results obtained by
the time-dependent Schr\"odinger equation, and underlying physics were well understood by
the adiabatic approximation based on the envelope Hamiltonian.
In this paper, we further explore two more 
aspects of the detachment dynamics, which were not done in our previous work. 
First, we find out features of both a {\it spatial} and {\it temporal} interference
of photo electron wave packets in a photo absorption process. We conclude that
both the interference mechanisms are universal in ionization dynamics
in high-frequency regime. To our knowledge, it is first time that
both the interference mechanisms in high-frequency regime 
are extracted from the first principle. 
Second, we extract a pulse duration 
which maximize a yield of the non-adiabatic transition as a function of a pulse duration. 
It is shown that it becomes maximum when the pulse duration is comparable to a time-scale 
of an electron. 
\end{abstract}
\maketitle
\section{Introduction}
The latest experimental techniques of high order harmonic
generations can generate coherent light sources in soft
x-ray range \cite{CA}, and opened up the new realm of the research area
so-called high-frequency regime. Here the terminology
``high-frequency'' means that a photon energy is high enough
to ionize a ground state electron by single photon absorption.
Meanwhile, high-frequency regime has been intensively 
studied in theory more than three decades in terms of
the high-frequency Floquet theory (HFFT) for
monochromatic laser fields developed by Gavrila and Kaminski \cite{GK}. 
The HFFT is developed in the Kramers-Henneberger (KH) frame \cite{Henne}.
In the KH frame, an effect of the laser field is described by
an atomic potential quivering along a classical trajectory
of a free electron in the laser fields. This is called the KH 
potential. In high-frequency limit, where the single optical
cycle of the laser field is much shorter than the electron's time scale,
it was shown that all the Fourier components of the KH potential can be ignored except 
the zeroth component i.e. a time average of the KH potential \cite{GK}. 
This is often called the dressed potential.
All the other photon absorption/emission channels then can be treated
perturbatively for bound states of the dressed potential even these 
amplitudes are comparable with the
dressed potential. 

After the foundation of the HFFT, a lot of
literatures have been involved to study ionization dynamics
in high-frequency regime. One of the most striking physical 
phenomenon in high-frequency regime is the stabilization, 
which an ionization rate begins to decrease 
for an intensity higher than a certain critical value, first found by
Pont and Gavrila \cite{PG}. A great number of literatures had been devoted 
to understand this counter-intuitive phenomenon. You et al. \cite{YM}
showed that the stabilization stems from a {\it spatial}
interference of photo electron wave packets launched at two turning points
of the classical electron in laser fields. For small values of its quiver
amplitude, the interference is constructive because they are produced
almost the same positions in space. However, this picture turns
into destructive for a quiver amplitude larger than a certain critical value.
This is the origin of the stabilization.

The HFFT has given us interesting physical insights
in high-frequency regime, but it can only be applied for monochromatic
laser fields i.e. {\it infinite} pulse duration.
However, laser pulses of attosecond time scale has become available in the latest 
experiments as mentioned above. These unprecedented laser pulses 
will be employed to study light-matter interactions 
in extremely short time scale in high-frequency regime, where 
effects of a time-varying envelope function of a pulse is expected
to play important roles. Therefore, it is highly desirable to develop
theoretical methods in high-frequency regime to adequately treat
{\it finite} pulse duration beyond the HFFT. 

Under such circumstances, we developed the 
envelope Hamiltonian to treat photo ionization dynamics in high-frequency regime
in our previous work \cite{TS}. 
Photo electron amplitudes were analytically derived in the framework of the
adiabatic approximation based on the envelope Hamiltonian. The procedures follows the HFFT
i.e. we realize a dressed potential and treat photon absorption/emission 
channels perturbatively but the time-dependency of a pulse envelope is explicitly remained.
Thus we also obtain the photo electron amplitudes for a non-adiabatic transition 
induced by the time-dependent envelope function which does not show up in 
the original HFFT.
The capability of the envelope Hamiltonian and the adiabatic approximation
were demonstrated in \cite{TS} utilizing a simple model in one dimension in the stabilization 
regime. It was shown that the results obtained by the full time-dependent Schr\"odinger
equation (TDSE) calculations were accurately reconstructed by the TDSE for the envelope Hamiltonian. 

In this paper, we further explore ionization dynamics in high-frequency regime
working on two subjects utilizing the envelope Hamiltonian. First, we revisit 
the oscillating substructure in photon absorption peaks in high-frequency regime,
which has been recently studied by several groups \cite{TT, DC, YF, YMad}.
In \cite{TT}, they found the oscillating structure in the stabilization regime, 
and reconstructed it taking into account a {\it spatial} and {\it temporal} 
interference of photo electron wave packets. In the stabilization regime, 
ionization probability as a function of time has two peaks before and after a peak 
intensity due to the {\it spatial} interference of photo electron wave packets.
This means that a photo electron wave packet of a certain energy
is created in a rising and falling part of a pulse, and they 
provoke the {\it temporal} interference whose phase difference is given by 
different moments of their birth in time. However, their formulas were obtained 
in an empirical way incorporating a quasi static
picture into the HFFT. On the other hand, in \cite{DC, YF, YMad}, they only addressed
the {\it temporal} interference although their theoretical approach was based on
the first principle. We consider that they did not need to take into account the {\it spatial}
interference because their maximum quiver amplitude
of a free electron in their pulse was below a threshold for the emergence of the stabilization.
In this paper, to our knowledge, we find out for the first time both the signature 
of the {\it spatial} and {\it temporal} interference 
in the formula obtained from the first principle. 

The second subject in this paper is to extract an optimal pulse duration to maximize 
a yield of the non-adiabatic transition. It is found that the yield as a function of a pulse 
duration has a maximum at a certain pulse duration in our previous work \cite{TS}. 
We find out a formula to predict
the peak position, and show that the yield becomes
maximum when the pulse duration is close to a time scale of an electron.
As far as we know, these two subjects have not been explored yet in high-frequency regime
due to the lack of appropriate theoretical frameworks which can take into account
time-varying envelope functions. So, we strongly believe that these are worth to gain
further insights for ionization dynamics in high-frequency regime. 

This paper is thus organized as follows. In Sec.~\ref{sec:theory}, we introduce
our theoretical methods. We briefly summarize the formulations in \cite{TS} for a
case of one dimension to refer them in later. In Sec.~\ref{sec:theory}A, B, and C,
we derive the envelope Hamiltonian in the KH frame employing a normalized classical
trajectory. In Sec.~\ref{sec:theory}D, we derive photo electron amplitudes
for the photon adsorption/emission processes and non-adiabatic transition. 
After considering on the non-adiabatic transition in Sec.~\ref{sec:nat},
we extract the evidence of the {\it spatial} and {\it temporal} interference in the formula
of the photon absorption spectrum with the aid of the saddle point method in Sec.~\ref{sec:interference}. 
In Sec.~\ref{sec:results}, we study a detachment dynamics of a model negative ion
in high-frequency regime to demonstrate our theory. In Sec.~\ref{sec:ATI}, we revisit
the oscillating structure in photon absorption peaks to confirm
that our theory is consistent with previously known results \cite{TT, DC, YF, YMad}.
In Sec.~\ref{sec:optimal}, we find out an optimal pulse duration to maximize
a yield of non-adiabatic transition. In Sec.~\ref{sec:conclusion}, we conclude
the paper with future perspectives. Atomic units are used thorough out the paper.

\section{Theoretical methods}
\label{sec:theory}
In this section, we summarize our theoretical method \cite{TS}
in one dimension to refer them in later. 
\subsection{TDSE in the Kramers-Henneberger frame}
The time-dependent Schr\"odinger equation in one dimension reads,
\begin{equation}
H(t)|\Psi(t)\rangle=i\frac{\partial}{\partial t}|\Psi(t) \rangle.
\label{eq:tdse}
\end{equation}
The Hamiltonian $H(t)$ within dipole approximation 
in the Kramers-Henneberger (KH) frame is given by
\begin{equation}
H(t)=-\frac{1}{2}\frac{\partial^2}{\partial x^2}+V(x+x_\omega(t)),
\label{eq:fullH}
\end{equation}
where $V(x)$ is an atomic potential.
The function $x_\omega(t)$ represents a classical trajectory of
a free electron in a laser pulse $F(t)$,
\begin{equation}
\label{eq:pulse}
\frac{d^2 x_\omega}{dt^2}=-F(t).
\end{equation}
The laser pulse $F(t)$ satisfies
\begin{subequations}
\begin{eqnarray}
\lim_{t \to \pm \infty} F(t)=0,\\
\int_{-\infty}^\infty F(t^\prime)dt^\prime=0.
\end{eqnarray}
\label{eq:physicalpulse}
\end{subequations}
In the KH frame, external fields are described by 
quivering motions of the atomic potential 
along the classical trajectory Eq.~(\ref{eq:pulse}),
and this potential is called the KH potential.

\subsection{Normalized classical trajectory}
We define the classical trajectory $x_\omega(t)$ in Eq.~(\ref{eq:fullH}) as
\begin{equation}
x_\omega(t)=\alpha(t)\cos(\omega t+\delta).
\label{eq:xomega}
\end{equation}
Then the pulse is given by Eq.~(\ref{eq:pulse}).
The function $\alpha(t)$ is the envelope of the pulse
given by
\begin{equation}
\alpha(t)=\alpha_0e^{-a\left(\frac{t}{T}\right)^2}
\label{eq:alpha}
\end{equation}
where $a=2\log(2)$ so that a pulse duration $T$
is defined by the full width of half maximum (FWHM) of $F^2(t)$.
The constant $\alpha_0$ is given by
\begin{equation}
\alpha_0=\frac{F_0}{\omega^2+\frac{2a}{T^2}}
\label{eq:alpha0}
\end{equation}
so that the peak field amplitude becomes $F(0)=F_0\cos\delta$.
Note that our pulse $F(t)$ defined in this way
satisfies the conditions Eqs.~(\ref{eq:physicalpulse}).
In this paper, we consider values of a photon energy $\omega$ 
much higher than an ionization potential of an ground state, 
\begin{equation}
\omega \gg I_p.
\label{eq:hfcondition}
\end{equation} 
\subsection{The envelope Hamiltonian}
For a given value of the pulse envelope $\alpha(t)$, Eq.~(\ref{eq:alpha}), 
we introduce the function $V_n(x,t)$,
\begin{equation}
V_n(x,t)=\frac{1}{T_\omega}
\int_0^{T_\omega} V(x+\alpha(t)\cos(\omega t^\prime+\delta))
e^{in\omega t^\prime}dt^\prime.
\label{eq:vn}
\end{equation}
Let us consider how many functions $V_n(x,t)$
are needed to reconstruct the KH potential. First, we
consider a short pulse limit $T \to 0$. In this case, 
the value of the envelope Eq.~(\ref{eq:alpha}) is very small; see Eq.~(\ref{eq:alpha0}). 
So, considering the Taylor expansion of the function $V_n(x,t)$
for $n=0,~\pm 1~$ and $\pm 2$ up to the order of $\alpha^2(t)$,
\begin{subequations}
\begin{eqnarray}
\label{eq:V0app}
V_0(x,t) &\approx& V(x)+\frac{1}{4}V^{\prime \prime}(x)\alpha^2(t),\\
\label{eq:V1app}
V_{\pm 1}(x,t) &\approx& \frac{1}{2}V^{\prime}(x)\alpha(t)e^{\mp i\delta}, \\
\label{eq:V2app}
V_{\pm 2}(x,t) &\approx& \frac{1}{8}V^{\prime \prime}(x)
\alpha^2(t)e^{\mp 2i\delta}.
\end{eqnarray}
Here the prime represents spatial derivative.
We then obtain
\begin{equation}
\sum_{n=-2}^2 V_n(x,t)e^{-in\omega t} 
=V(x+x_\omega(t))+O(\alpha^3(t)).
\end{equation}
Second, we consider a long pulse limit $T \to \infty$. In this case,
the envelope function $\alpha(t)$ Eq.~(\ref{eq:alpha}) varies slowly in time
since a lot of optical cycles are contained in the pulse.
So, we can define the momentary Fourier expansion of the KH potential 
for a given time $t$ using the function $V_n(x,t)$ Eq.~(\ref{eq:vn}),
\begin{equation}
V(x+x_\omega(t))=\sum_{n=-\infty}^\infty V_n(x,t)e^{-in\omega t}.
\label{eq:fourier}
\end{equation}
\end{subequations}
This expression is exact for the long pulse limit $T \to \infty$. 
Since we work on high-frequency regime characterized by the inequality
Eq.~(\ref{eq:hfcondition}), it is enough to only consider the above
summation $|n| \leq 2$. 

Having confirmed that the KH potential can be accurately
approximated using a few terms of the function $V_n(x,t)$
for both the opposite time scale $T \to 0$ and $T \to \infty$, 
we introduce the envelope Hamiltonian $H_{\rm env}(t)$
\begin{equation}
\label{eq:Henv}
H_{\rm env}(t)=H_0(t)+U(x,t),
\end{equation}
where $H_0(t)$ and $U(x,t)$ are defined by
\begin{subequations}
\begin{eqnarray}
\label{eq:H0}
H_0(t)&=&-\frac{1}{2}\frac{\partial^2}{\partial x^2}+V_0(x,t),\\
\label{eq:uxt}
U(x,t)&=&\sum_{n=\pm 1,\pm 2} V_n(x,t)e^{-in\omega t}
\end{eqnarray}
\end{subequations}
and we consider the TDSE for $H_{\rm env}(t)$,
\begin{equation}
\label{eq:envtdse}
H_{\rm env}(t)|\psi(t)\rangle=i\frac{\partial}{\partial t}|\psi(t) \rangle.
\end{equation}
We call this equation the envelope TDSE. In the above, we separated
the function $V_0(x,t)$ from Eq.~(\ref{eq:fourier}), and used it to construct 
the quantity $H_0(t)$, Eq.~(\ref{eq:H0}). The function $H_0(t)$
in Eq.~(\ref{eq:H0}) reduces to the atomic Hamiltonian for $t \to \pm \infty$.
So, it is considered that the quantity $H_0(t)$ represents a 
distorted Hamiltonian of the electron during the action of the pulse.
It is often called the dressed Hamiltonian.
The equivalent Hamiltonian for the case of monochromatic fields
was also considered by Henneberger in \cite{Henne}. In the paper, he pointed out
that a fast convergence of a photo ionization cross section 
in a strong field can be achieved in perturbation theory based 
on $H_0(t)$, since all the orders of the field amplitude $F_0$
are included in its eigen function and eigen energy. 

\subsection{Adiabatic approximation for photo ionization}
\label{sec:adiabatic}
In this subsection, we implement an adiabatic approximation
for photo electron amplitudes. We derive
photo electron amplitudes in one-dimension 
based on the envelope Hamiltonian, Eq.~(\ref{eq:Henv}), 
to utilize them in later sections.
The full derivations with an arbitrary number of bound states in three dimension 
are found in \cite{TS}.
Let $|0(t) \rangle$ and $|k,t\rangle$ be a ground state 
and scattering state of the dressed Hamiltonian $H_0(t)$ 
Eq.~(\ref{eq:H0}),
\begin{subequations}
\begin{eqnarray}
H_0(t)|0(t)\rangle &=&E_0(t)|0(t) \rangle, \\
H_0(t)|k,t \rangle&=&E|k,t \rangle,
\end{eqnarray}
\end{subequations}
where
\begin{equation}
E=\frac{k^2}{2}.
\end{equation}
The orthogonality is
\begin{subequations}
\begin{eqnarray}
\langle 0(t)| 0(t) \rangle&=&1,\\
\langle k,t|k^\prime,t \rangle&=&2\pi \delta(k-k^\prime).
\end{eqnarray}
\end{subequations}
Employing them, we expand the solution of the envelope 
TDSE Eq.~(\ref{eq:envtdse}), 
\begin{eqnarray}
|\psi(t) \rangle=e^{-i\chi(t)}
\left[
C_0(t)|0(t) \rangle e^{-i\int_{-\infty}^t E_0(t^\prime)dt^\prime} 
+\int_{-\infty}^\infty C_{k^\prime}(t)|k^\prime,t \rangle e^{-iE^\prime t}
\frac{dk^\prime}{2\pi}
\right],
\label{eq:aexpansion}
\end{eqnarray}
where $E^\prime=k^{\prime 2}/2$ and the coefficients $C_0(t)$ and $C_k(t)$ represent the ground
state population and photo electron amplitude of momentum $k$ for a certain time $t$,
respectively. The phase $\chi(t)$ is given by
\begin{equation}
\chi(t)=\int_{-\infty}^t 
\langle 0(t^\prime)|
U(x,t^\prime)-i\frac{\partial}{\partial t^\prime}
|0(t^\prime) \rangle
dt^\prime,
\end{equation}
so that coupled differential equations for the coefficients
$C_0(t)$ and $C_k(t)$ become simple. Let us introduce the notation
\begin{equation}
\label{eq:Phin}
\Phi_n(t)=-\int_{-\infty}^t E_0(t^\prime)dt^\prime-n\omega t+Et.
\end{equation}
Substituting the expansion
Eq.~(\ref{eq:aexpansion}) into the envelope TDSE Eq.~(\ref{eq:envtdse}),
we obtain
\begin{subequations}
\label{eq:volterra}
\begin{eqnarray}
i\frac{d C_0}{dt}&=&e^{i\int_{-\infty}^t E_0(t^\prime)dt^\prime} 
\int_{-\infty}^\infty Q_{0k^\prime}(t)C_{k^\prime}(t)e^{-iE^\prime t}
\frac{dk^\prime}{2\pi},\\
i\frac{\partial C_k}{\partial t}&=& Q_{k0}(t)C_0(t)
e^{i\Phi_0(t^\prime)} 
+\int_{-\infty}^\infty Q_{kk^\prime}(t)C_{k^\prime}(t^\prime)e^{-i(E^\prime-E)t}\frac{dk^\prime}{2\pi},
\end{eqnarray}
where 
\begin{equation}
\label{eq:qmat}
Q_{k0}(t)=\langle k,t|
U(x,t)-i\frac{\partial}{\partial t}
|0(t) \rangle,~{\rm etc}.
\end{equation}
Using the Hellmann-Feynman theorem, the quantity $Q_{k0}(t)$ becomes
\begin{equation}
Q_{k0}(t)=\langle k,t|U(x,t)+i\frac{\dot V_0(x,t)}{E-E_0(t)}|0(t) \rangle,
\end{equation}
\end{subequations}
where $\dot V_0$ is time derivative of $V_0$.
Eqs.~(\ref{eq:volterra}) can be solved perturbatively.
Let us assume the zeroth order solution as $C_0^{(0)}(t)=1$ and 
$C_{k}^{(0)}(t)=0$. 
Then the first order solution is given by
\begin{subequations}
\label{eq:1storder}
\begin{eqnarray}
C_0^{(1)}(t)&=&1, 
\label{eq:c0}\\
C_k^{(1)}(t)&=&-i\int_{-\infty}^t 
Q_{k0}(t^\prime)e^{i\Phi_0(t^\prime)} dt^\prime
\label{eq:c1}
\end{eqnarray}
\label{eq:c1-2}
\end{subequations}
The photo electron spectrum $dp/dk$ is thus approximated as,
\begin{equation}
\frac{dp}{dk} \simeq \left|\lim_{t \to \infty} C_k^{(1)}(t)\right|^2
=\left|\sum_{n=-2}^2 C^{(1)}_{n\omega}(k)\right|^2.
\label{eq:dpdk}
\end{equation}
The function $C_{n\omega}^{(1)}(k)$ is defined by
\begin{subequations}
\begin{equation}
\label{eq:pea}
C_{n\omega}^{(1)}(k)=\int_{-\infty}^\infty
M_{n\omega}(k,t)
e^{i\Phi_n(t)}dt,
\end{equation}
where
\begin{equation}
\label{eq:mkt}
M_{n\omega}(k,t)=
\left \{
\begin{array}{cr}
\langle k,t|i\frac{\dot V_0(x,t)}{E-E_0(t)}|0(t)\rangle & (n=0), \\
\langle k,t|V_n(x,t)|0(t)\rangle & (n\neq 0). 
\end{array}
\right.
\end{equation}
\end{subequations}
The total ionization yield is given by
\begin{equation}
P_{\rm ion}=\int_{-\infty}^{\infty} 
\left|\sum_{n=-2}^2C^{(1)}_{n\omega}(k)\right|^2
\frac{dk}{2\pi}.
\label{eq:pion}
\end{equation}
The ionization yield by each channel is given by
\begin{equation}
P_{n}=\int_{-\infty}^\infty |C^{(1)}_{n\omega}(k)|^2\frac{dk}{2\pi}.
\label{eq:pn}
\end{equation}
\subsection{Non-adiabatic transition}
\label{sec:nat}
The total photo electron amplitude Eq.~(\ref{eq:pea}) consists of 
two different kinds of physical processes. The first one is
given by $n=0$ in Eq.~(\ref{eq:mkt}) which represents
the non-adiabatic transitions to the continuum
induced by the time-dependency of the ground state 
$|0(t) \rangle$. Therefore, one speculates that 
the following TDSE is responsible for the non-adiabatic transition,
\begin{equation}
H_0(t)|\psi_0(t) \rangle
=i\frac{\partial}{\partial t}|\psi_0(t) \rangle.
\label{eq:nattdse}
\end{equation}
The TDSE Eq.~(\ref{eq:nattdse}) accurately approximates
the full TDSE Eq.~(\ref{eq:tdse}) if the
contributions of multi photon absorption/emission 
to the total ionization yield are negligibly small.  
As far as we know, the TDSE Eq.~(\ref{eq:nattdse})
is first realized in \cite{BO} to study the non-adiabatic transition
between bound states in one dimension. In the paper, they predicted that
an electron is ionized with low energy. 
In Eq.~(\ref{eq:mkt}) for $n=0$, the integrand
has a large amplitude around the energy $E=E_0(t)<0$,
and tends to be zero as $E \to \infty$. So  
it is expected that the non-adiabatic transition ionizes an electron with
low energies. The generation of a slow electron in high-frequency regime
was first confirmed in \cite{FS}. In the paper, it was found
when the electron subjects to a square-shaped pulse i.e. sudden jump between
a field free and dressed ground state is responsible for the emergence.
However the mechanism is different to the non-adiabatic transition here since
time-derivative of the dressed potential $V_0(x,t)$ in Eq.~(\ref{eq:H0})
cannot be defined at the moment of the sudden ramp of the pulse. 
Later, the emergence of the slow electron in the context of the non-adiabatic transition 
was obtained in \cite{TT2}, whose spectrum for a long pulse limit is studied in terms of 
the adiabatic approximations to the transitions to the continuum \cite{Tol}.
More recently, the slow electron was also found in the study of
an above threshold ionization spectrum of a carbon atom in hard x-ray regime \cite{TK}.
They explained its emergence by the Raman type process i.e. 
a single photon absorption followed by a single photon emission.
We consider that this corresponds to the lowest order approximation 
to the non-adiabatic transition, Eq.~(\ref{eq:V0app}), which is 
also second order with respect to $\alpha(t)$, Eq.~(\ref{eq:alpha}).

\subsection{Spatial and temporal interference of photo electron wave packets}
\label{sec:interference}
\begin{figure}
\centerline{\includegraphics[scale=0.5]{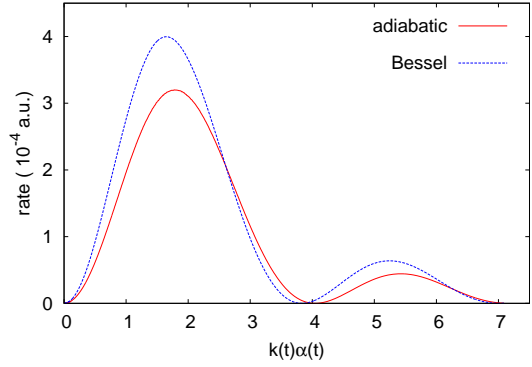}}
\caption{\label{fig:gamma1}
A solid line shows the single photon absorption rate as a function of $|k(t)|\alpha(t)$,
given by Eq.~(\ref{eq:Gamman}) for $n=1$.
Here, $k(t)=\sqrt{2(E_0(t)+\omega)}$ is a momentum of an ionized electron,
and $\alpha(t)$ the envelope function of our pulse defined in Eq.~(\ref{eq:alpha}). 
A dot line shows an asymptotic result of Eq.~(\ref{eq:Gamman}) for $\omega \to \infty$; 
see Eq.~(\ref{eq:mat_asym}). An oscillating feature of the rate comes from the Bessel function. 
}
\end{figure}

Another contribution in Eq.~(\ref{eq:pea}) is defined by Eq.~(\ref{eq:mkt})
 with $n \neq 0$. For positive (negative) values of $n$, this function represents 
the photo electron amplitude by $n$photon absorption (emission).
The stationary phase condition is given by
\begin{equation}
\label{eq:sphase}
E=E_0(t)+n\omega.
\end{equation}
This equation has two solutions $t=t_{\mp}~(t_+=-t_-)$ in the rising
and falling part of the pulse, respectively. Taking into account
these two solutions, the spectrum is approximated to
\begin{subequations}
\label{eq:sp}
\begin{equation}
\label{eq:dp1dk}
|C_{n\omega}^{(1)}(k)|^2\approx \frac{4 \Gamma_n(t_+)}{|\dot{E_0}(t_+)|}
\cos^2\left(\theta(t_+)+\frac{\pi}{4}\right),
\end{equation}
where
\begin{eqnarray}
\label{eq:Gamman} 
\Gamma_{n}(t_+)&=&2\pi |M_{n\omega}(k,t_+)|^2,\\
k(t)&=&\pm \sqrt{2(E_0(t)+n\omega)}.
\end{eqnarray}
The function $\Gamma_n(t)$ represents the $n$photon ionization rate
at a given time $t$, and the quantity $k$ the momentum of the ionized electron.
The sign of $k$ corresponds to an electron ionizing to
the right~($+$) and left~($-$) direction, respectively. 
In the derivation of Eq.~(\ref{eq:dp1dk}), the relations of $\Gamma_n(t_-)=\Gamma_n(t_+)$ and
$\dot{E_0}(t_-)=|\dot{E_0}(t_+)|$ are used; Our envelope function $\alpha(t)$,
Eq.~(\ref{eq:alpha}) is symmetric with respect to $t=0$.
The function $\theta(t_+)$ is given by
\begin{equation}
\label{eq:theta}
\theta(t_+)=-\int_{0}^{t_+}E_0(t)dt+E_0(t_+)t_+.
\end{equation}
\end{subequations}
It is found in Eq.~(\ref{eq:dp1dk}) that two different interference mechanisms contribute to
the formation of the photon absorption spectrum. 
The first mechanism is a {\it spatial} interference found in the
$n$photon ionization rate $\Gamma_n(t)$, which is extracted by 
employing the derivation by Pont \cite{Pont} regarding the envelope
function $\alpha(t)$ as an adiabatic variable.
Then the rate $\Gamma_n(t)$ in high-frequency limit 
$\omega \to \infty$ can be approximated as
\begin{subequations}
\begin{equation}
\label{eq:mat_asym}
\Gamma_n(t)\approx
|A(k(t))|^2|\varphi_0(\alpha(t))|^2 J_n^2(|k(t)|\alpha(t)),
\end{equation}
where
\begin{equation}
A(k(t))=\int_{-\infty}^\infty V(x)e^{ik(t)x} dx, 
\end{equation}
and
\begin{equation}
\varphi_0(-\alpha(t))=\frac{1}{T_\omega}\int_0^{T_\omega}
\phi_0(-\alpha(t)\cos(\omega t^\prime+\delta))dt^\prime,
\end{equation}
\end{subequations}
and the function $J_n(z)$ represents the Bessel function of $n$th order.
In the derivation of Eq.~(\ref{eq:mat_asym}), the scattering 
state $|k,t \rangle$ is approximated as
\begin{equation}
|k,t \rangle \approx e^{ik(t)x},
\end{equation}
and also the following function as
\begin{eqnarray}
\phi_0(-\alpha(t)\cos(\omega t^\prime+\delta))
&=&\sum_{m=-\infty}^\infty \varphi_m(-\alpha(t))e^{-im\omega t^\prime} \nonumber \\
&\approx& \varphi_0(-\alpha(t)).
\end{eqnarray}
We would emphasize that the functional form of Eq.~(\ref{eq:mat_asym})
is universal i.e. it does not depend on dimensionality and number of bound states
in an atomic potential. Indeed, Pont's derivation was done for hydrogen atom in
three-dimension \cite{Pont}. The key quantity to understand Eq.~(\ref{eq:mat_asym})
is the Bessel function which oscillates as a function of $|k(t)|\alpha(t)$.
This is shown in Fig.~\ref{fig:gamma1}. This figure compares the single photon absorption
rate Eq.~(\ref{eq:Gamman}) with $n=1$ and its asymptotic form Eq.~(\ref{eq:mat_asym})
for a model potential of H$^-$ Eq.~(\ref{eq:h-pot}). 
It is found that the oscillation comes from the Bessel function in Eq.~(\ref{eq:mat_asym}).
The argument $|k(t)|\alpha(t)$ represents the phase difference between two photo electron wave packets 
of the same energy $E=\frac{1}{2}k^2(t)$ produced by $n$photon ionization channel $V_n(x,t)$ at the
positions of $x=\pm \alpha(t)$ which are the two turning points of the classical 
free electron at a certain time $t$ in the pulse. The 
interference of these two photo electron wave packets are 
constructive for small values of the argument $|k(t)|\alpha(t)$. 
However, if the argument can exceed 
a certain threshold during the action of the pulse,
the picture of the interference turns into destructive, which happens 
around $|k(t)|\alpha(t) \approx 1.8$. 
This is the emergence of the stabilization in high-frequency regime \cite{YM}. The 
oscillating feature of the Bessel function in Eq.~(\ref{eq:mat_asym})
represents that constructive and destructive interference appear one after the another
as a function of $|k(t)|\alpha(t)$. Such a behavior was also found in
numerical calculations in \cite{YC1992}.

The second interference mechanism in Eq.~(\ref{eq:dp1dk}) is the {\it temporal} interference
imprinted in $\cos$ function. 
The creation of the photo electron wave packet of the energy $E$ by the {\it spatial} interference
takes place twice i.e. in the rising and falling part of the pulse, respectively.
They interfere with the phase difference given by Eq.~(\ref{eq:theta}), which represents
the difference of the accumulation of the dynamical phase between them. The formula quite similar to
Eq.~(\ref{eq:sp}) was found in the study of the oscillating substructure
in photon absorption peaks in \cite{TT}. The difference is that our formula 
does not take into account the depletion of the ground state since Eq.~(\ref{eq:dp1dk})
is obtained from the first order solution to Eq.~(\ref{eq:volterra}); see Eqs.~(\ref{eq:1storder}).
However, the formula in \cite{TT} was obtained in an empirical manner introducing a
quasi static picture into the HFFT. In doing so, they obtained the single photon absorption rate
as a function of time which exhibited clearly separated two peaks before and after the peak
field amplitude of the pulse, which is the emergence of the stabilization i.e. 
the signature of the {\it spatial} interference in destructive way. 
This gave them the idea which two photo electron 
wave packets produced by the {\it spatial} interference 
in the rising and falling part of the pulse cause the {\it temporal} interference.
However, as far as we know, it is first time to obtain the formula
Eq.~(\ref{eq:dp1dk}) from the first principle which can capture the signatures of 
both the {\it spatial} and {\it temporal} interference in the spectrum.
After the findings in \cite{TT}, several groups had also found the {\it temporal}
interference in high-frequency regime for hydrogen atom in \cite{DC}, and
also for hydrogen molecular ion \cite{YF, YMad}. However, in these studies, only
the {\it temporal} interference was discussed since their pulse parameters are off
the stabilization regime.

\subsection{Numerical implementations}
In this paper, we mainly study a photo detachment of 
hydrogen negative ion H$^-$ in one dimension with single active 
electron approximation. The electron's potential is modeled by
\cite{PT}
\begin{subequations}
\begin{eqnarray}
V(x)=-D\frac{\exp[-\sqrt{x^2+a^2}]}{\sqrt{x^2+b^2}}, \\
D=24.856,~~a=4,~~b=6.27. 
\end{eqnarray}
\label{eq:h-pot}
\end{subequations}
This potential supports only one bound state $E_0=-0.0277$.
We employ the Siegert state expansion method in the KH frame, 
previously developed in \cite{TT} to solve the full TDSE Eq.~(\ref{eq:tdse})
and the envelope TDSE Eq.~(\ref{eq:envtdse}), and also Eq.~(\ref{eq:nattdse}).

\section{Results}
\label{sec:results}
\subsection{Revisit of the oscillating substructure in photon absorption peaks}
\label{sec:ATI}
\begin{figure}
\centerline{\includegraphics[scale=0.75]{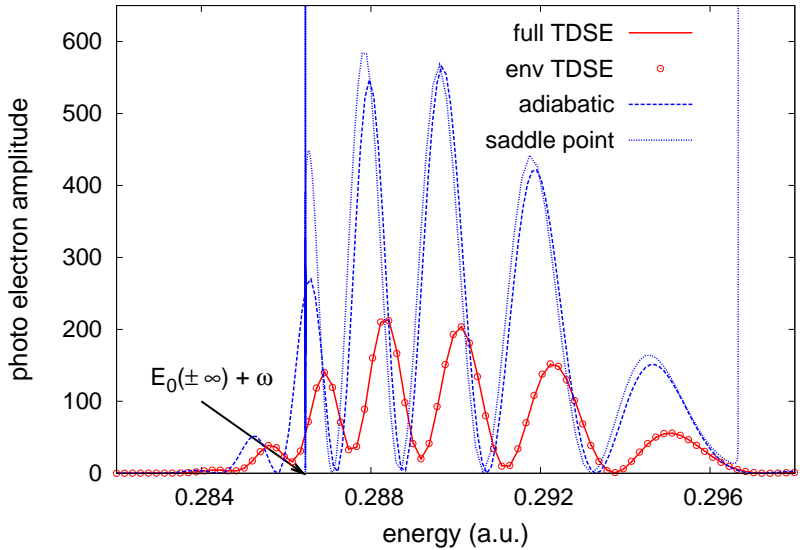}}
\caption{\label{fig:interference} 
Photo electron spectrum for the laser parameters $F_0=0.5$, $\omega=\pi/10$,
and $T=2000$ near a photo peak $E=E_0(\pm \infty)+\omega$ indicated by an
arrow. 
The solid (green) and broken (red) lines are obtained by
the envelope TDSE, Eq.~(\ref{eq:envtdse}) and the saddle point method
Eq.~(\ref{eq:dp1dk}).
}
\end{figure}

Fig.~\ref{fig:interference} shows a photo electron spectrum
near a position of a photo peak $E=E_0(\pm \infty)+\omega$ for a set of 
laser parameters $F_0=0.5$, $\omega=\pi/10$ and $T=2000$. The
solid line (red) and blank circles (red) represent the result
of the full TDSE Eq.~(\ref{eq:tdse}) and envelope TDSE Eq.~(\ref{eq:envtdse}), respectively.
It is clearly shown that the envelope TDSE perfectly reproduces
the full TDSE result. The overall structure is blue-shifted with respect to
the photo peak at $E=E_0(\pm \infty)+\omega$ since the ground
state energy $E_0(t)$ becomes shallower during the action of the pulse.
The dotted (blue) and broken (blue) lines
represent the result of the adiabatic approximation Eq.~(\ref{eq:pea}) for $n=1$
and saddle point method Eq.~(\ref{eq:dp1dk}). 
The amplitudes of them are quite overestimated compared to the full TDSE calculation
because the depletion of the ground state is ignored in the adiabatic
approximation; Eq.~(\ref{eq:pea}) is the first order solution to Eq.~(\ref{eq:volterra}).
And the phase shift of the interference structure is found for these results
with respect to the full TDSE result Eq.~(\ref{eq:tdse}). This is also
due to the ignore of the depletion in the adiabatic approximation;
the photo electron amplitude produced in the falling part of the pulse is 
largely overestimated. So, the relative phase of it with respect to that
produced in the rising part of the pulse deviates from the exact calculation. 
The divergence around $E \approx E_0(\pm \infty)+\omega=0.2864$ 
and $E \approx E_0(0)+\omega=0.2965$ in the saddle
point method are seen because time derivative of the ground state energy vanishes;
see Eq.~(\ref{eq:dp1dk}). The former comes from the rising and falling edge of
the pulse where the Stark dressing to the ground state is very small, and the latter
at the peak intensity of the pulse. 
Since the saddle points $t_-$ and $t_+$ coalesce at the peak intensity, the
divergence around the high energy edge can be removed by the uniform approximation 
\cite{Berry}. The procedure is given in Appendix \ref{app:uniform}. 
We consider that the oscillating substructure can be understood 
very well with the adiabatic approximation and saddle point method. 
Therefore, we thus conclude that the oscillating 
structure in photon absorption peak in high-frequency regime
is formed by the {\it spatial} and {\it temporal} interference of
photo electron wave packets discussed in Sec.~\ref{sec:interference}.

\subsection{Time scale of non-adiabatic transition}
\label{sec:optimal}
In this subsection, we solve Eq.~(\ref{eq:nattdse}) for several model potentials 
to study the non-adiabatic transition. In the previous work \cite{TS}, it was found that
a yield of non-adiabatic transition has a maximum as a function of a pulse duration. 
One may consider that this is a mathematical artifact due to the normalization factor
$\alpha_0$, Eq.~(\ref{eq:alpha0}), for our classical trajectory Eq.~(\ref{eq:xomega}).
For a limit of a pulse duration $T \to 0$, the yield vanishes because $\alpha_0$ becomes
zero. And the yield also vanishes for a limit of $T \to \infty$ because the dressed
potential $V_0(x,t)$ in Eq.~(\ref{eq:H0}) varies infinitely slowly in time. Therefore,
it is no wonder that a maximum can be found in between. However, a position of the maximum 
can be found in a region where the function $\alpha_0$ is almost converged to
its asymptotic value $F_0/\omega^2$ for the limit $T \to \infty$, and the position 
is far away from a region where the function $\alpha_0$ rapidly converges to zero.
Hence, the maximum has a physical origin rather than the mathematical artifact.

We consider small values of $\alpha_0$ to facilitate ourselves to derive formulas 
in perturbation theory to extract physics of the non-adiabatic transition. 
Up to the second order of $\alpha^2(t)$, a photo electron amplitude for the non-adiabatic transition
$C^{(1)}_{0\omega}(k)$, Eq.~(\ref{eq:pea}) for $n=0$, is reduced to,
\begin{subequations}
\begin{eqnarray}
\label{eq:cnatapp}
C_{0\omega}^{(1)}(k) &\approx& D(k)f(E),\\
\label{eq:m2}
D(k)&=&\langle k,t=-\infty|V^{\prime \prime}(x)|0(-\infty) \rangle,\\
f(E)&=&-\frac{a}{T^2}\frac{1}{E-E_0(\pm \infty)}
\int_{-\infty}^\infty t\alpha^2(t)
e^{i\Phi_0(t)}dt.
\end{eqnarray}
\end{subequations}
The function $\Phi_0(t)$ is given in Eq.~(\ref{eq:Phin}).
For simplicity, we ignore the Stark shift in the function $f(E)$.
Then the function $f(E)$ is reduced to 
\begin{subequations} 
\begin{equation}
f(E) \approx 
-i\sqrt{\frac{\pi}{32a}}\alpha_0^2T
e^{-\frac{T^2}{8a}(E-E_0^{(0)})^2}.
\label{eq:f2eapp}
\end{equation}
\end{subequations}
Therefore, we obtain the approximated spectrum of the non-adiabatic transition,
\begin{equation}
|C_{0\omega}^{(1)}(k)|^2= 
\frac{\pi}{32a}|D(k)|^2\alpha_0^4T^2
e^{-\frac{T^2}{4a}(E-E_0)^2},
\label{eq:natapprox}
\end{equation}
Next, we derive the formula for the detachment yield. 
To implement this, we consider an ansatz for the functional form of $|D(k)|^2$,
\begin{subequations}
\begin{equation}
\label{eq:m2ansatz}
|D(k)|^2=c_1k^2e^{-c_2\frac{k^2}{4aI_p}}.
\end{equation}
\end{subequations}
where $I_p=|E_0(\pm \infty)|$. The detachment probability $P_0(T)$ as a function of
a pulse duration $T$ is calculated substituting Eq.~(\ref{eq:m2ansatz}) into (\ref{eq:natapprox}), 
and integrating over $k$,
\begin{subequations}
\begin{eqnarray}
P_0(T)&=&\int_{-\infty}^\infty |C_{0\omega}^{(0)}(k)|^2\frac{dk}{2\pi} \nonumber \\
&=&\frac{c_1}{2^{\frac{15}{4}}a^{\frac{1}{4}}}
\alpha_0^4\sqrt{T}\xi^{\frac{3}{2}}
e^{\xi^2-\frac{(I_pT)^2}{4a}} 
\left[K_{\frac{3}{4}}(\xi^2)-K_\frac{1}{4}(\xi^2)\right],
\label{eq:p0app}
\end{eqnarray}
where the $\xi$ is defined as
\begin{equation}
\xi=\frac{1}{\sqrt{8a}}\frac{(I_pT)^2+c_2}{I_pT}.
\label{eq:xi}
\end{equation}
\end{subequations}
The derivation is found in Appendix \ref{app:derivepnat}.
In Appendix \ref{app:c2}, it is clarified that the constant $c_2$
is related to the curvature of the atomic potential. 

Before calculating a position of a maximum of Eq.~(\ref{eq:p0app}),
we take a limit of $\omega \to \infty$ for the formula. In doing so, the function
$\alpha_0$ in Eq.~(\ref{eq:p0app}) sharply increases from 0, and quickly converges to 
its asymptotic value $F_0/\omega^2$ as $T$ increases. Then the maximum of Eq.~(\ref{eq:p0app}) 
takes place where the value of $\alpha_0$ is enough converged. It is thus guaranteed that 
the occurrence of the maximum of Eq.~(\ref{eq:p0app}) does not stem from our normalization
factor $\alpha_0$, Eq.~(\ref{eq:alpha0}), for our normalized classical trajectory Eq.~(\ref{eq:xomega}). Now we attempt to analytically extract an
optimal pulse duration from Eq.~(\ref{eq:p0app}). 
We approximate the Bessel functions of fractional order $K_{\nu}(z)~(\nu=3/4~{\rm and}~1/4)$ 
in Eq.~(\ref{eq:p0app}) using its asymptotic form for large arguments,
\begin{equation}
K_{\nu}(z) \sim \sqrt{\frac{\pi}{2z}}e^{-z}\left[1+\frac{4\nu^2-1}{8z}\right].
\label{eq:Kasym}
\end{equation}
This approximation becomes valid for $T \to 0$ and $T \to \infty$ since the argument
$\xi$, Eq.~(\ref{eq:xi}), diverges for these limit.
The yield of the non-adiabatic ionization then becomes,
\begin{equation}
\label{eq:pnatasym}
P_0 \sim \frac{c_1\sqrt{\pi a}\alpha_0^4}{16\sqrt{I_p}}
\frac{(I_pT)^2}{[(I_pT)^2+c_2]^{\frac{3}{2}}}e^{-\frac{(I_pT)^2}{4a}}.
\end{equation}
The position of maximum is found solving $dP_0/d(I_pT)=0$.
Then we obtain,
\begin{subequations}
\begin{equation}
(I_pT)^4+(2a+c_2)(I_pT)^2-4ac_2=0,
\end{equation}
The solution is
\begin{equation}
\label{eq:Tasym}
I_pT=\sqrt{\frac{1}{2}
\left[
-2a-c_2+\sqrt{4a^2+20ac_2+c_2^2}
\right]}.
\end{equation}
\end{subequations}

\begin{figure}
\begin{tabular}{cc}
\includegraphics[scale=0.5]{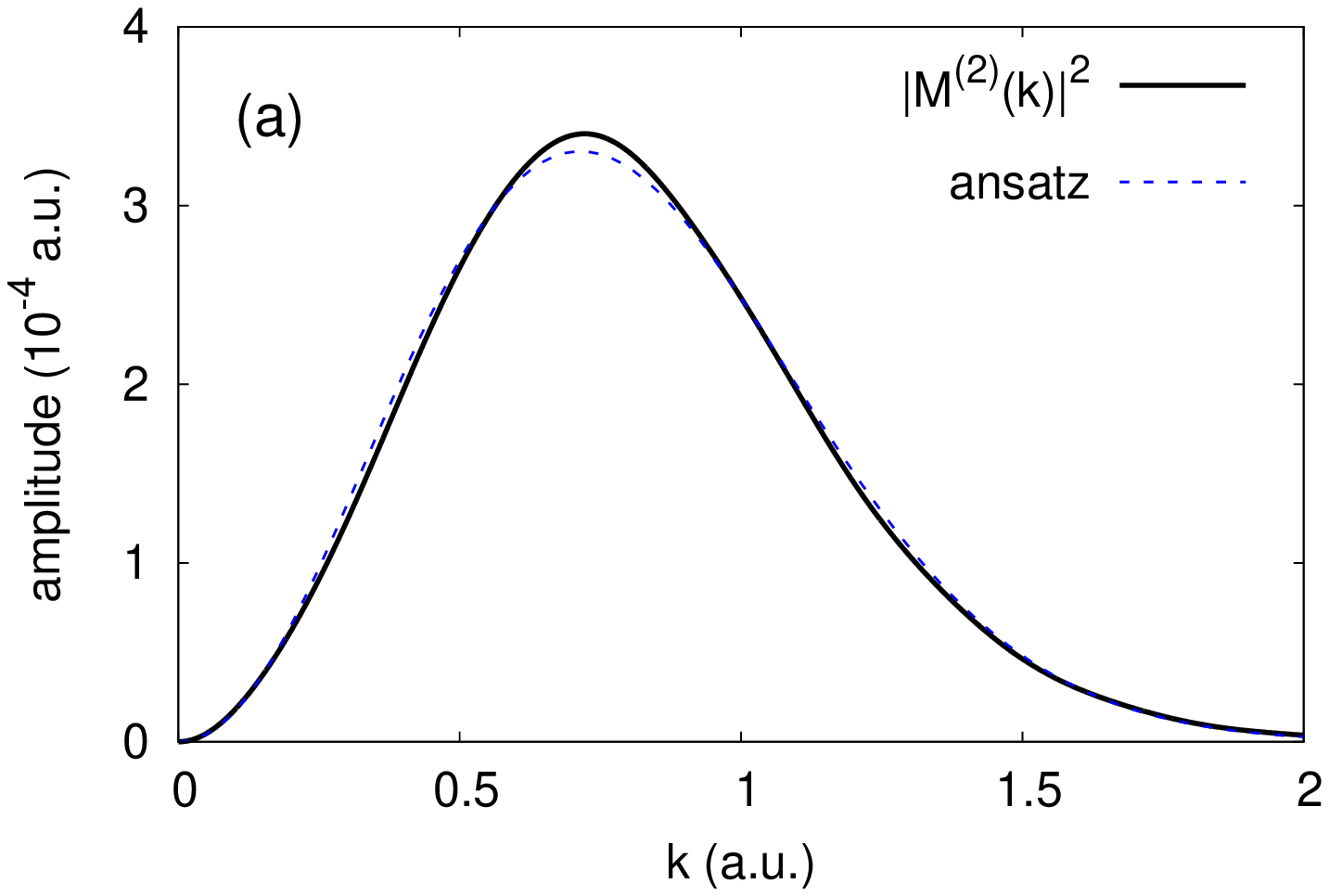} & \includegraphics[scale=0.5]{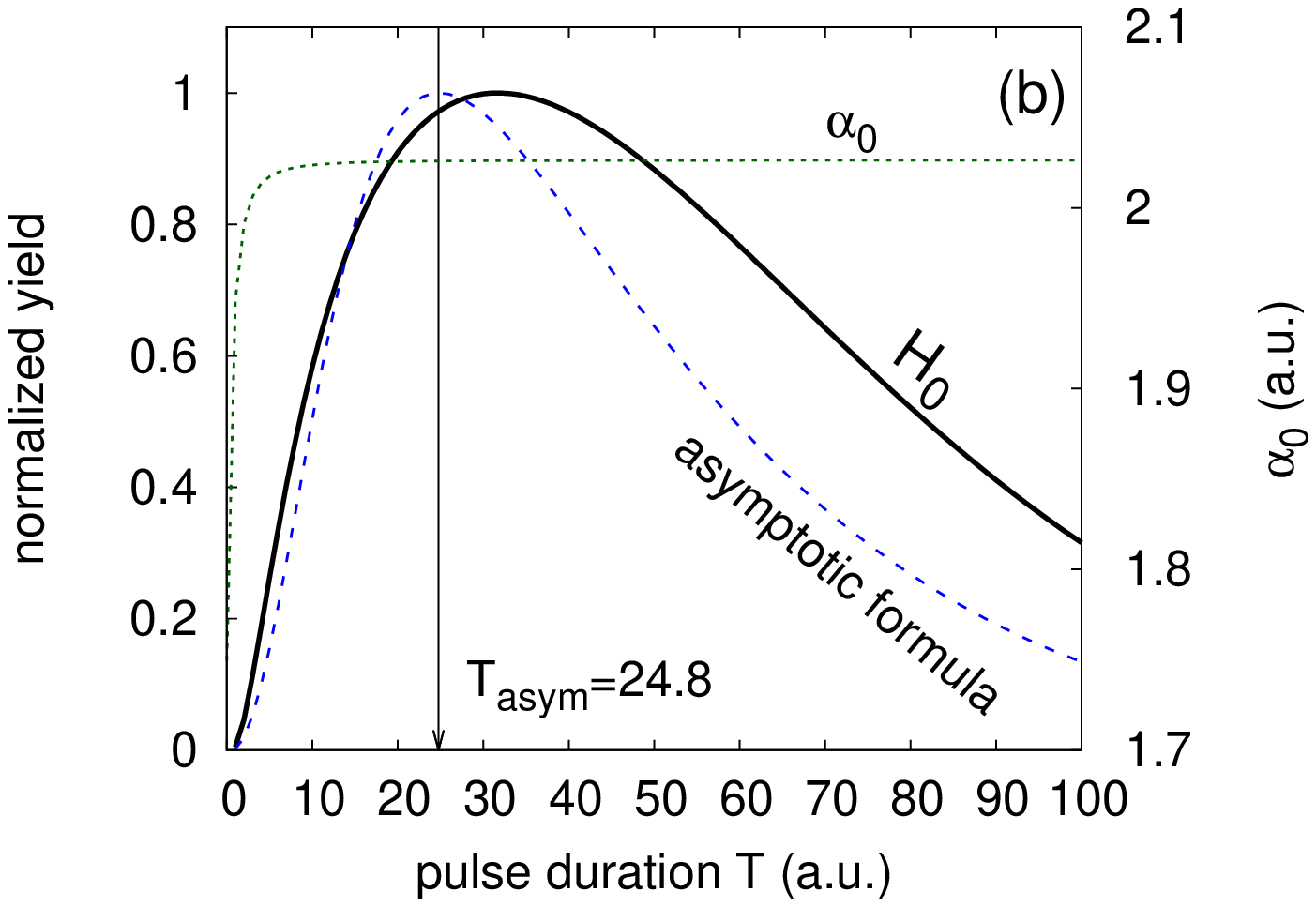}
\end{tabular}
\caption{\label{fig:3} 
(a) Comparison between the numerically calculated matrix element $M^{(2)}(k)$, Eq.~(\ref{eq:m2}), and
its fitting by the ansatz Eq.~(\ref{eq:m2ansatz}). 
(b) The solid line (black) shows the detachment yield by the non-adiabatic transitions 
for our model, Eq.~(\ref{eq:h-pot}), as a function of pulse duration 
obtained solving Eq.~(\ref{eq:nattdse}) with $\omega=4\pi/5$ and $F_0=12.8$. 
The broken line (blue) is obtained
using the asymptotic formula Eq.~(\ref{eq:pnatasym}) for $\omega \to \infty$ under 
the ansatz Eq.~(\ref{eq:m2ansatz}). The maximum yields of these results are normalized 
to unity to clearly compare the positions of the peaks. 
The dot line shows the normalization factor $\alpha_0$ of our envelope function Eq.~(\ref{eq:alpha0}). 
}
\end{figure}

We first demonstrate the result Eq.~(\ref{eq:Tasym}) for a model potential of H$^-$ Eq.~(\ref{eq:h-pot}).
The parameters $c_1$ and $c_2$ for ansatz Eq.~(\ref{eq:m2ansatz}) are found using fitting
procedures, we obtain $c_1=1.76 \times 10^{-3}$ and $c_2=0.301$. 
The result of the fitting is shown in Fig.~\ref{fig:3}(a).
The solid (black) and dotted (blue) lines show the result obtained by 
Eqs.~(\ref{eq:m2}) and (\ref{eq:m2ansatz}), respectively. We find that both the results
agree very well. Next, we consider the high-frequency limit $\omega \to \infty$ to apply Eq.~(\ref{eq:Tasym}).
To this end, we introduce a scaling of $F_0$ and $\omega$ as 
\begin{equation}
F_0 \to m^2 F_0~~~{\rm and}~~~\omega \to m\omega.
\label{eq:scale}
\end{equation}
to keep a ratio $F_0/\omega^2$ being a constant, where $F_0=0.2$ and $\omega=\pi/10$.
For the values of $m=1,2,4$ and $8$,
we carried out solving Eq.~(\ref{eq:nattdse}). 
We terminated our calculations at $m=8$
since a peak position of the yield of the non-adiabatic transition 
it not significantly different comparing the result
for $m=4$. This means that $T$ dependency in the function $\alpha_0$, 
Eq.~(\ref{eq:alpha0}) is washed out around a region where the peak position locates.
Fig.~\ref{fig:3}(b) shows the detachment yield of the non-adiabatic transition.
The solid (black) and broken (blue) line are obtained by the 
TDSE Eq.~(\ref{eq:nattdse}) and the asymptotic formula Eq.~(\ref{eq:pnatasym}), respectively. The height
of these lines are normalized to unity to clearly compare the peak positions.
Substituting the value of the ionization potential $I_p=0.0277$ and $c_2=0.301$ to Eq.~(\ref{eq:Tasym}), 
the expected optimal pulse duration is found to be $T_{\rm asym}=24.8$. This is about $20\%$ off the exact 
value $T=31$ shown in Fig.~\ref{fig:3}. We consider that this is reasonable in the rough approximations.

\subsubsection{Effect of atomic structure on optimal pulse duration}
\begin{figure}
\begin{tabular}{cc}
\includegraphics[scale=0.5]{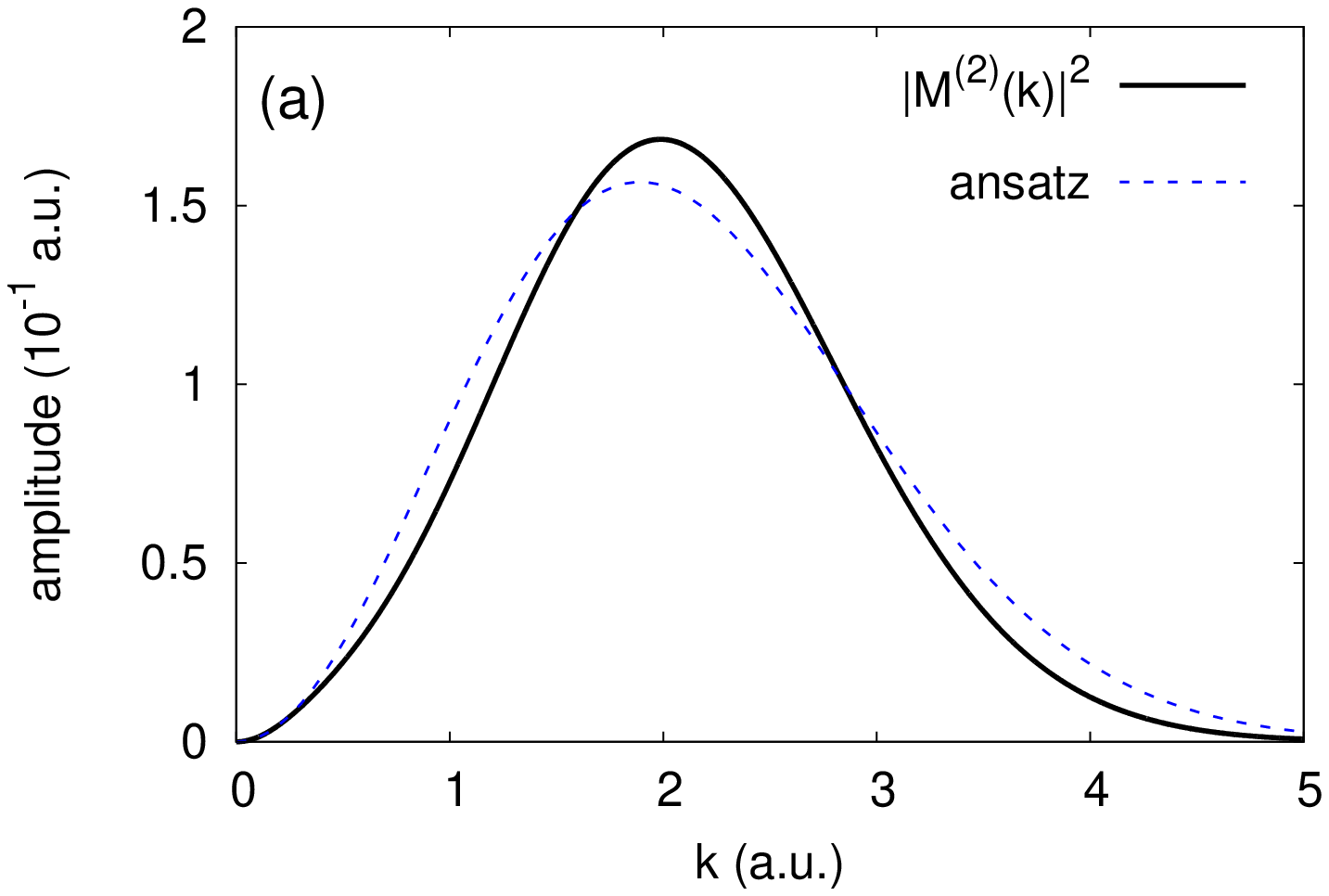} & \includegraphics[scale=0.5]{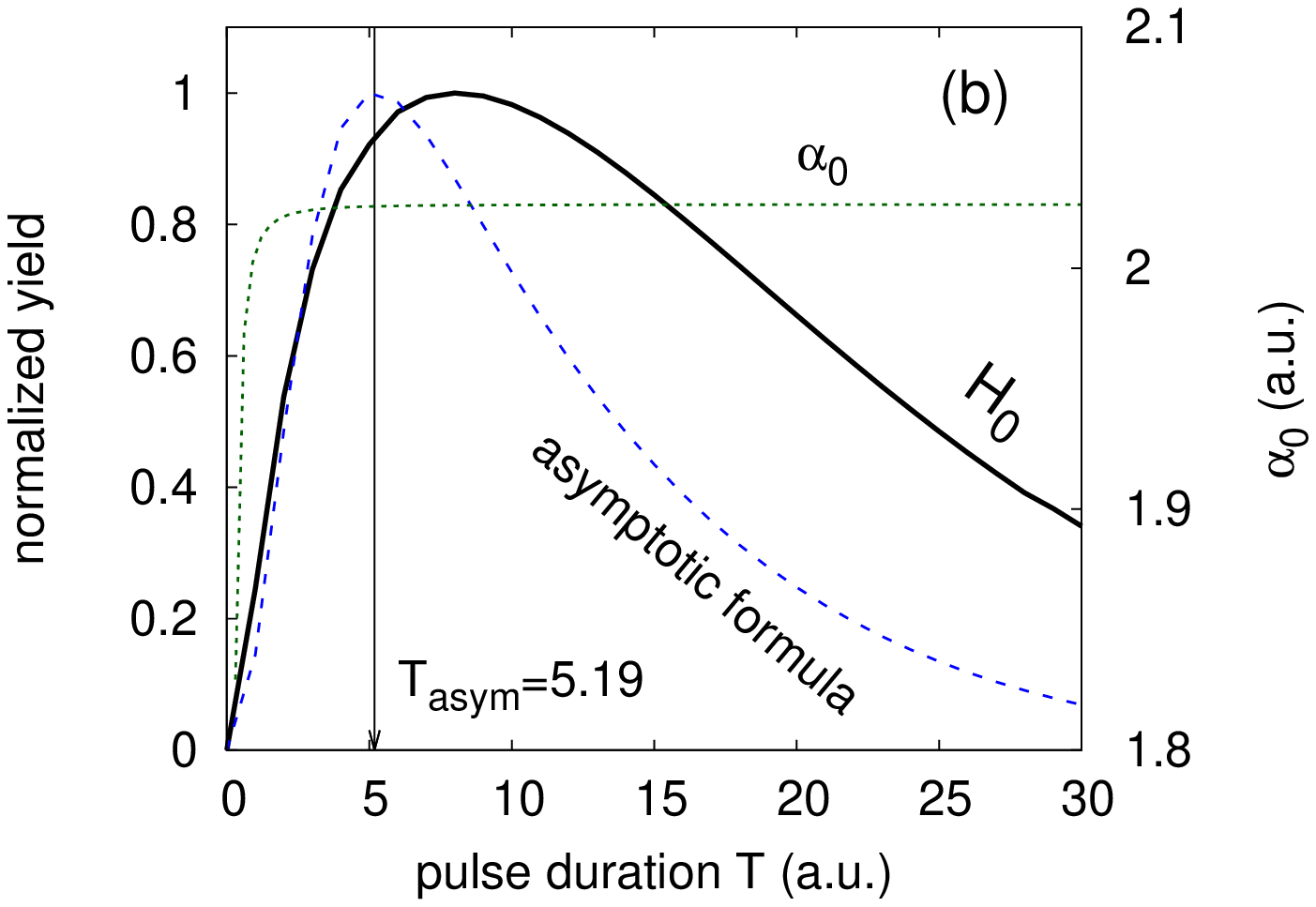}  
\end{tabular}
\caption{\label{fig:4} 
The same as Fig.~\ref{fig:3} but another atomic potential Eq.~(\ref{eq:gpot})
with the depth $W_0$ and the width $\sigma$ referred as case I in Table \ref{tab:1}.
}
\end{figure}

\begin{figure}
\begin{tabular}{cc}
\includegraphics[scale=0.5]{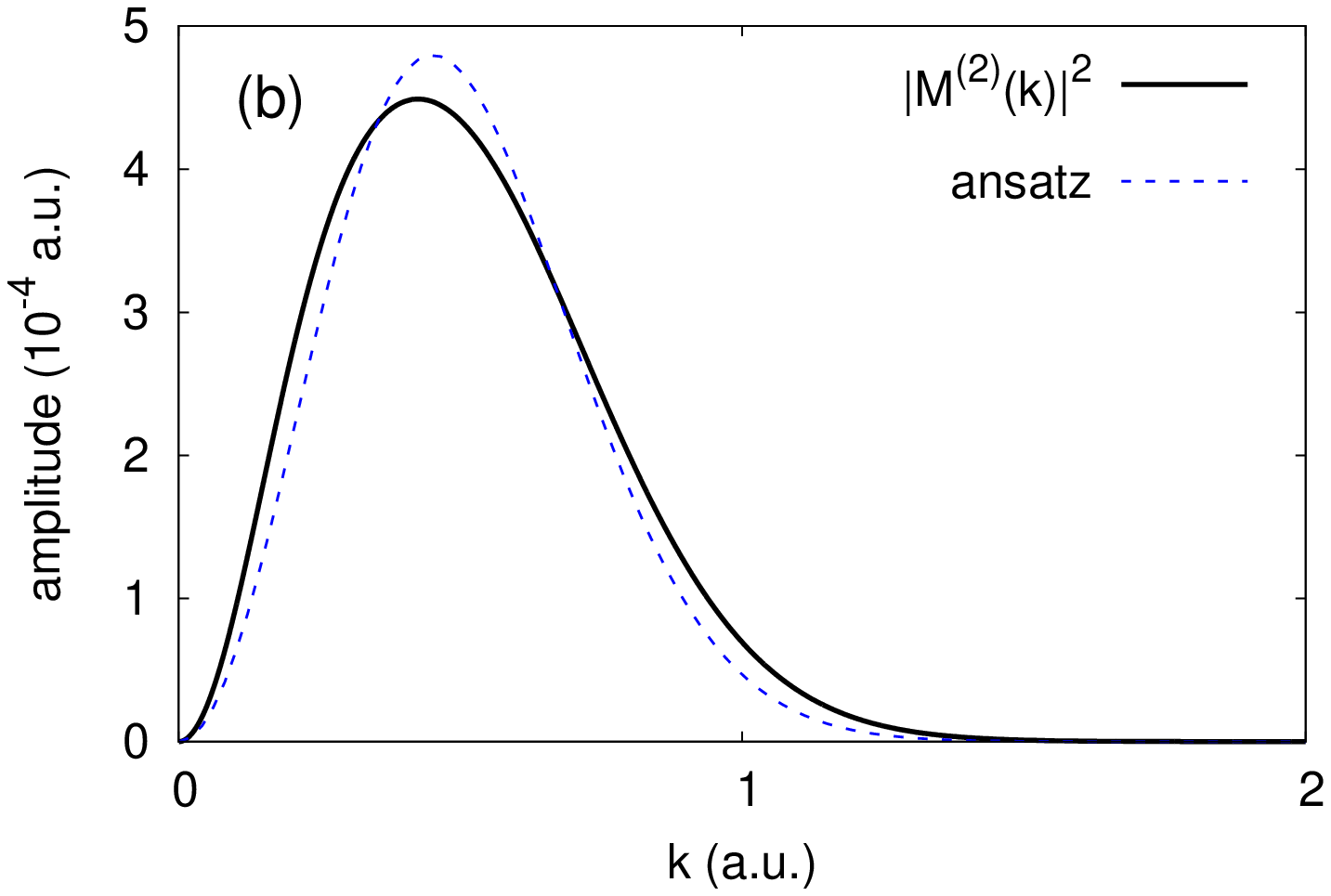} & \includegraphics[scale=0.5]{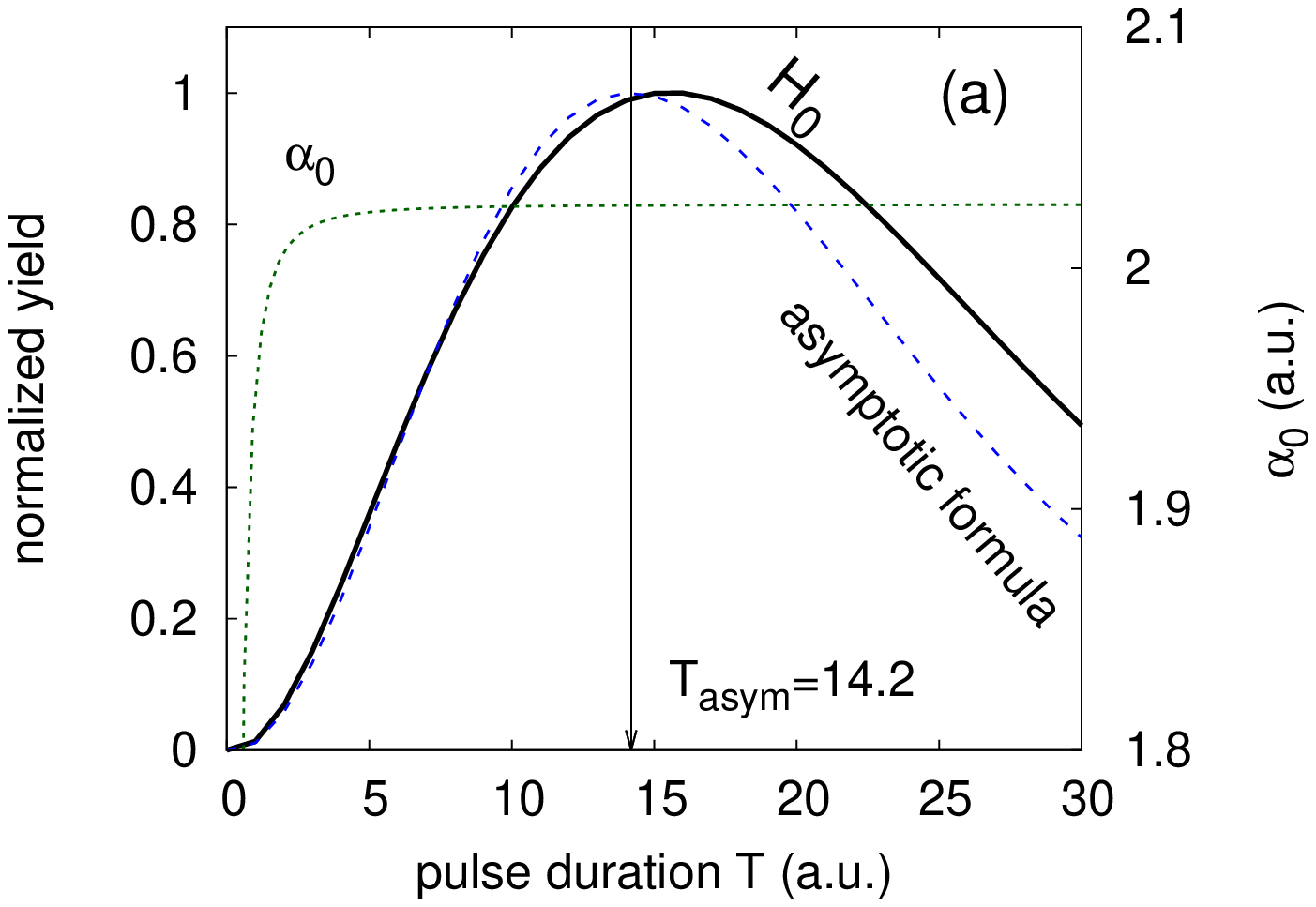} 
\end{tabular}
\caption{\label{fig:5} 
The same as Fig.~\ref{fig:3} but another atomic potential Eq.~(\ref{eq:gpot})
with the depth $W_0$ and the width $\sigma$ refereed as case II in Table \ref{tab:1}.
}
\end{figure}

We find that Eq.~(\ref{eq:Tasym}) depends on the constant $c_2$, which is related
to the curvature of the atomic potential $V(x)$. In harmonic approximation
of the atomic potential, the curvature gives us a ground state energy; see Appendix \ref{app:c2}. 
To see effects of atomic structure on the optimal pulse duration, let us consider another atomic potential
\begin{equation}
\label{eq:gpot}
W(x)=-W_0e^{-(x/\sigma)^2},
\end{equation}
with different combinations of the parameters $W_0$ and $\sigma$
summarized in the table \ref{tab:1}, which are referred as case I and II,
respectively. The case I~(II) represents a deep and narrow (shallow and wide)
atomic potential. These combinations are chosen so that the field free ground 
state energy becomes $E_0(\pm \infty)=-0.1$. 
The values of $c_1$ and $c_2$ for the ansatz Eq.~(\ref{eq:m2ansatz}) are summarized
in Table \ref{tab:1}. We repeated the scaling procedure Eq.~(\ref{eq:scale})
to reach high-frequency limit $\omega \to \infty$. In the case I (II), a peak position
of a maximum yield of the non-adiabatic transition is converged for $m=32~(m=16)$.
Substituting the parameter
$c_2$ into Eq.~(\ref{eq:Tasym}), which are given in table \ref{tab:1}, 
we obtain
\begin{subequations}
\label{eq:optimal-g}
\begin{eqnarray}
{\rm case~I}~~~I_pT \approx 0.519~\rightarrow~T_{\rm asym}&=&5.19,\\
{\rm case~II}~~~I_pT \approx 1.42~\rightarrow~T_{\rm asym}&=&14.2.
\end{eqnarray}
\end{subequations}
These values are also found in Table \ref{tab:1}.
Results are shown in Fig~\ref{fig:4} and \ref{fig:5} for the case I and II, respectively,
in a manner of Fig.~\ref{fig:3}. The quality of the asymptotic expansion
of the modified Bessel function Eq.~(\ref{eq:Kasym})
near the origin for case II is better than the case I due to the bigger value of $c_2$;
see Table \ref{tab:1}. So, we obtain the better result for the position of the maximum yield
in the case II than the case I.

It was shown in Fig.~4 of \cite{Tol} that the yield of the non-adiabatic
transition has a maximum for a certain value
of a pulse duration, and it was estimated using $I_pT \approx 1$, which
is equivalent to Eq.~(\ref{eq:Tasym}) with the right hand side being unity.
However, the right hand side of Eq.~(\ref{eq:Tasym}) depends on not only
binding energies but also curvatures of target potentials as shown in this 
demonstration. 
\begin{table}
\setlength{\tabcolsep}{9pt}
\begin{tabular}{|c|cc|cc|c|}
\hline
        & $W_0$ & $\sigma$ & $c_1$ & $c_2$ & $T_{\rm asym}$  \\
\hline
case I  & 0.345 & 1 & 0.119   & 0.155 & 5.19 \\ 
case II & 0.159 & 4 & 0.00641 & 2.73  & 14.2 \\
\hline
\end{tabular}
\caption{\label{tab:1} Different combination of the parameters $W_0$ and $\sigma$ 
for the atomic potential Eq.~(\ref{eq:gpot}) so that a field free 
ground state energy becomes $E_0(\pm \infty)=-0.1$. The parameters $c_1$ 
and $c_2$ are used for the ansatz Eq.~(\ref{eq:m2ansatz}) to fit
the matrix element $M^{(2)}(k)$, Eq.~(\ref{eq:m2}).
The value of $T_{\rm asym}$ is a predicted peak position of the detachment
yield by the non-adiabatic transition Eq.~(\ref{eq:p0app}) 
using asymptotic formula Eq.~(\ref{eq:Tasym}).
}
\end{table}

\section{conclusion}
\label{sec:conclusion}
Following the previous work \cite{TS}, 
we further explored a detachment dynamics of a model negative ion 
in high-frequency regime. 
We revisited the interference substructures in photon adsorption peaks
in an adiabatic approximation based on the envelope Hamiltonian.
The adiabatic approximation clarified that two different interference mechanisms
are responsible for its emergence.
The first mechanism is the {\it spatial} interference. At a certain time in a rising
part of a pulse, two photo electron wave packets are launched at two
turning points of a classical electron in the pulse. An interference of them
create net amount of a photo electron wave packet. This {\it spatial} interference
is repeated in the falling part of the pulse. Then these photo electron wave packets 
produced in different moments in time cause {\it temporal} interfere. We confirmed that the adiabatic
approximation can well reproduce the oscillating substructure obtained from
the full time-dependent Schr\"odinger equation (TDSE). The interference substructure was previously found 
in \cite{TT}, and recently the same mechanism was confirmed for hydrogen atom in \cite{DC, YMad}. 
We showed that our theory is consistent with these known results. 
In \cite{TT}, they predicted the coexistence of the {\it spatial} and {\it temporal} interference.
However, their formulation was based on an empirical approach bringing a quasi static
picture into the high-frequency Floquet theory \cite{GK}. So, to our best
knowledge, it is first time to find out both the interference mechanisms
derived from the first principle.  

We also extracted an optimal pulse duration to maximize a detachment yield
by non-adiabatic transition. We clarified that the yield is maximized
for a pulse duration close to time scale of non-adiabatic transition,
roughly estimated by Eq.~(\ref{eq:Tasym}). 

Our demonstrations have been done utilizing short range potential, although our
formulation does not depend on dimensionality and properties of atomic 
potential \cite{TS}. Further studies in three-dimension in a real atomic system 
will be worked out in future.

\section{Acknowledgements}
K. T. would thank for Profs. Ulf Saalmann and Jan M. Rost
for the discussions to improve the manuscript.

\appendix
\section{Derivation of Eq.~(\ref{eq:p0app})}
\label{app:derivepnat}
Substituting Eq.~(\ref{eq:m2ansatz}) into (\ref{eq:natapprox}), 
and integrating over $k$, the ionization yield of 
the non-adiabatic transition $P_0(T)$ is written as
\begin{subequations}
\begin{equation}
P_0(T)=\int_{-\infty}^\infty |C_{0\omega}(k)|^2\frac{dk}{2\pi}
=\frac{c_1\alpha_0^4\beta}{4} e^{-\frac{(I_pT)^2}{4a}}
\int_{-\infty}^\infty 
k^2e^{-\beta k^4-\gamma k^2}dk,
\end{equation}
where
\begin{eqnarray}
\beta&=&\frac{T^2}{16a},\\
\gamma&=&\frac{I_pT^2}{4a}+\frac{c_2}{4aI_p}.
\end{eqnarray}
\end{subequations}
The integral can be written using 
the fractional order $\nu$ of the
modified Bessel function $I_\nu(z)$ of first kind,
\begin{subequations}
\begin{eqnarray}
\int_{-\infty}^\infty
x^2 e^{-\beta x^4-\gamma x^2}dx
=\frac{\pi e^{\frac{\gamma^2}{8\beta}}}{8\sqrt{2}\beta^{\frac{3}{2}}\sqrt{\gamma}}
\left[
-\gamma^2I_{-\frac{1}{4}}\left(\frac{\gamma^2}{8\beta}\right)
+(4\beta+\gamma^2)I_{\frac{1}{4}}\left(\frac{\gamma^2}{8\beta}\right)\right. \nonumber \\
\left.+\gamma^2\left\{
-I_{\frac{3}{4}}\left(\frac{\gamma^2}{8\beta}\right)
+I_{\frac{5}{4}}\left(\frac{\gamma^2}{8\beta}\right)
\right\} 
\right].
\end{eqnarray}
Using the properties of $I_\nu(z)$ \cite{nist},
\begin{eqnarray}
I_{\nu-1}(z)-I_{\nu+1}(z)=\frac{2\nu}{z}I_\nu(z),\\
K_\nu(z)=\frac{\pi}{2}\frac{I_{-\nu}(z)-I_\nu(z)}{\sin(\nu \pi)},
\end{eqnarray} 
where $K_\nu(z)$ is the modified Bessel function of second kind,
the integral can be simplified to
\begin{equation}
\int_{-\infty}^\infty
x^2 e^{-\beta x^4-\gamma x^2}dx=
\frac{1}{8}\left(\frac{\gamma}{\beta}\right)^{\frac{3}{2}}
e^{\frac{\gamma^2}{8\beta}}
\left[K_{\frac{3}{4}}\left(\frac{\gamma^2}{8\beta}\right)
-K_{\frac{1}{4}}\left(\frac{\gamma^2}{8\beta}\right)\right],
\end{equation}
\end{subequations}
Therefore, we obtain
\begin{equation}
P_0(T)=\frac{c_1}{2^{\frac{15}{4}}a^{\frac{1}{4}}}
\alpha_0^4\sqrt{T}\xi^{\frac{3}{2}}
e^{\xi^2-\frac{(I_pT)^2}{4a}}
\left[
K_{\frac{3}{4}}(\xi^2)-K_\frac{1}{4}(\xi^2)
\right],
\end{equation}
where
\begin{equation}
\xi=\frac{1}{\sqrt{8a}}\frac{(I_pT)^2+c_2}{I_pT}.
\end{equation}

\section{Physical meaning of the constant $c_2$}
\label{app:c2}
The time-independent Schr\"odinger equation
for the ground state $\phi_0$ with the energy $E_0$
reads, 
\begin{equation}
\left[-\frac{1}{2}\frac{d^2}{dx^2}+V(x)\right]\phi_0(x)=E_0\phi_0(x),
\end{equation}
Let us consider the Taylor expansion of the atomic potential $V(x)$
around the origin up to the order of $x^2$, 
\begin{equation}
V(x) \approx V(0)+\frac{1}{2}\Omega^2 x^2,
\label{eq:vpotapp}
\end{equation}
where $\Omega^2=V^{\prime \prime}(0)$ represents the second derivative of atomic potential
$V(x)$ at origin.
In this approximation, the ground state $\phi_0^{(0)}(x)$ and its energy level
$E_0^{(0)}$ thus correspond to those of simple harmonic oscillator, which are 
given by
\begin{subequations}
\begin{eqnarray}
\label{eq:howf}
\phi_0^{(0)}(x)&=&\left(\frac{\Omega}{\pi}\right)^{\frac{1}{4}}e^{-\frac{\Omega}{2}x^2}, \\
\label{eq:hoene}
E_0^{(0)}&=&V(0)+\frac{\Omega}{2}.
\end{eqnarray}
\end{subequations}
In what follows, we assume the condition of $E_0 \approx E_0^{(0)}<0$. 
We exclude considering  the case of $E_0^{(0)}>0$ which can happen 
for steep atomic potentials.   

Now let us calculate the matrix element $M^{(2)}(k)$ Eq.~(\ref{eq:m2}). 
To this end, we consider the case of $k \gg 1$. Then the scattering
state can be replaced to $\langle k,t=\pm \infty|=e^{ikx}$. We approximate
the ground state wave function $|0,t=\pm \infty \rangle$ by Eq.~(\ref{eq:howf}). 
With these assumptions, substituting the expansion Eq.~(\ref{eq:vpotapp}) 
into Eq.~(\ref{eq:m2}),
\begin{eqnarray}
|M^{(2)}(k)|^2 &=&|\langle k,t=-\infty|V^{(2)}(x)|0,t=-\infty \rangle|^2 \nonumber \\
&\approx& \frac{\Omega^{\frac{9}{2}}}{\pi^{\frac{1}{2}}}
\left| \int_{-\infty}^\infty e^{-\frac{1}{2}\Omega x^2}e^{ikx}dx\right|^2 \nonumber \\
&=& 2\sqrt{\pi}\Omega^{\frac{7}{2}}e^{-\frac{k^2}{\Omega}}.
\end{eqnarray}
This is the asymptotic formula of Eq.~(\ref{eq:m2ansatz})
for $k \gg 1$. Therefore we obtain
\begin{equation}
\label{eq:c2app}
\frac{c_2}{4aI_p}=\frac{1}{\Omega},
~\rightarrow~c_2=\frac{4aI_p}{\Omega}.
\end{equation}
For the 1D model of H$^-$, Eq.~(\ref{eq:h-pot}) and case I in Table \ref{tab:1} for the Gaussian potential
Eq.~(\ref{eq:gpot}), the ground state energy
with harmonic approximation is bigger than $0$. So, the formulation in this appendix is
not applicable. For the case II in table \ref{tab:1},
$V^{(2)}(0)=0.02$ and $I_p=0.088$. On the other hand, the exact value is $I_p=0.1$. 
Then Eq.~(\ref{eq:c2app}) gives us $c_2=3.45$, while
the exact value shown in table~\ref{tab:1} is $c_2=2.72$.

\section{Uniform approximation}
\label{app:uniform}
The formulation here follows Berry \cite{Berry}. To implement the uniform approximation,
we introduce the mapping for Eq.~(\ref{eq:Phin})
\begin{equation}
\Phi_n(t)=-\int_{-\infty}^t E_0(t^\prime)dt^\prime-n\omega t+Et=\zeta y+\frac{y^3}{3}+X.
\label{eq:map}
\end{equation}
Let $t=t_{\pm}$ satisfies the stationary phase condition, 
\begin{equation}
\left(\frac{d\Phi}{dt}\right)_{t_\pm}=-E_0(t_\pm)-\omega+E=(\zeta+y^2)\left(\frac{dy}{dt}\right)_{t_\pm}=0,
\end{equation}
then these are mapped onto
\begin{equation}
y_{\pm}= \pm i\sqrt{\zeta}.
\label{eq:spy}
\end{equation}
Realising $y_+=-y_-$, the constant $X$ is given by
\begin{equation}
X=\frac{1}{2}\left[\Phi(t_-)+\Phi(t_+)\right].
\label{eq:consta}
\end{equation}
Substituting Eqs.~(\ref{eq:spy}) and (\ref{eq:consta}) into Eq.~(\ref{eq:map}), we obtain,
\begin{equation}
\zeta=\left[-\frac{3i}{2}\theta(t_+)\right]^{\frac{2}{3}}.
\end{equation}
Another mapping we need is,
\begin{equation}
\frac{dt}{dy}M_{n\omega}(k,t)=p+qy. 
\end{equation}
Substituting $y=y_{\pm}$ into this equation, the constants $p$ and $q$ are determined as
\begin{subequations}
\label{eq:pq}
\begin{eqnarray}
p&=&\frac{1}{2}\left[\left(\frac{dt}{dy}\right)_{y_+}+\left(\frac{dt}{dy}\right)_{y_-}\right]M_{n\omega}(k,t_+), \\
q&=&-\frac{i}{2\sqrt{\zeta}}\left[\left(\frac{dt}{dy}\right)_{y_+}-\left(\frac{dt}{dy}\right)_{y_-}\right]
M_{n\omega}(k,t_+)
\end{eqnarray}
\end{subequations}
Note that the matrix element of the  $n$photon absorption take the same value at $t=t_\pm$. 
To calculate the value of $dt/dy$, we twice differentiate Eq.~(\ref{eq:map}) by $y$,
\begin{equation}
2y=\left[-E_0(t)-\omega+E\right]\frac{d^2t}{dy^2}-\frac{dE_0}{dt}\left(\frac{dt}{dy}\right)^2
\end{equation}
Realising that $dt/dy$ is even function, substituting either $y=y_-$ or $y_+$, we obtain
\begin{subequations}
\begin{equation}
\label{eq:sol1}
\left(\frac{dt}{dy}\right)_{y_\pm}=\sqrt{\frac{-2i\zeta^{1/2}}{\dot{E_0}(t_{+})}},
\end{equation}
or
\begin{equation}
\label{eq:sol2}
\left(\frac{dt}{dy}\right)_{y_\pm}=-\sqrt{\frac{-2i\zeta^{1/2}}{\dot{E_0}(t_{+})}}.
\end{equation}
\end{subequations}
Substituting this into Eq.~(\ref{eq:pq}), we thus obtain 
\begin{subequations}
\begin{eqnarray}
\label{eq:pq2}
p&=&\pm\sqrt{\frac{-2i\zeta^{1/2}}{\dot{E_0}(t_{+})}}M_{n\omega}(k,t_+), \\
q&=&0.
\end{eqnarray}
\end{subequations}
The positive and negative sign of $p$ corresponds to the solution Eq.~(\ref{eq:sol1}) or (\ref{eq:sol2}),
respectively. Therefore, the photo electron amplitude for single photon absorption, Eq.~(\ref{eq:pea}) for $n=1$,
is given by
\begin{eqnarray}
\label{eq:uni}
C_{n\omega}^{(1)}(k)&=&e^{iX}\int_{-\infty}^{\infty}\frac{dt}{dy}M_{n\omega}(k,t) e^{i(\zeta y+\frac{y^3}{3})}dy \nonumber \\
&=&\pm 2\pi e^{iX}\sqrt{\frac{-2i\zeta^{1/2}}{E_0^\prime(t_{+})}}
M_{n\omega}(k,t_+) {\rm Ai}(\zeta) \nonumber \\
&=&\pm e^{iX}\sqrt{\frac{-4\theta(t_+)}{\dot{E_0}(t_+)}}M_{n\omega}(k,t_+)
J_{\frac{1}{3}}\left(-\theta(t_+)\right),
\end{eqnarray}
The function ${\rm Ai}(z)$ represents the Airy function. Here we used on the last line \cite{nist},
\begin{subequations}
\begin{eqnarray}
{\rm Ai}(z)&=&\frac{1}{\pi}\sqrt{\frac{z}{3}}K_{\frac{1}{3}}\left(\frac{2}{3}z^{\frac{3}{2}}\right), \\
K_{\nu}(z)&=&J_{\nu}(iz)
\end{eqnarray}
\end{subequations}
where the function $K_\nu(z)$ and $J_\nu(z)$ are the modified Bessel function of
fractional order $\nu$, and Bessel function of fractional order $\nu$, respectively. 
Substituting the asymptotic form of the matrix element for the $n$photon absorption 
Eq.~(\ref{eq:mat_asym}) into Eq.~(\ref{eq:uni}),
\begin{eqnarray}
\label{eq:pea_asym}
C_{n\omega}^{(1)}(k)&=&
\pm i^{n} e^{i(X-n\delta)}\sqrt{\frac{4\theta(t_+)}{\dot{E_0}(t_+)}}
A(k(t_+))\varphi_0(-\alpha(t_+)) \nonumber \\
&\times& J_n(|k(t_+)|\alpha(t_+)) J_{\frac{1}{3}}\left(-\theta(t_+)\right)
\end{eqnarray}
It is found that the spectrum is written using two Bessel functions.
The Bessel function of the integer order $n$ represents the {\it spatial}
interference, and the fractional order $1/3$ {\it temporal} interference.
It is easily shown by L'H\^opital's rule that the quantity 
$\theta(t_+)/\dot{E_0}(t)$ is order of $t_+^2$ at the vicinity of $t_+=0$.
Therefore, the result Eq.~(\ref{eq:pea_asym}) does not have the singularity
at $t_+=0$.

\end{document}